# Chat GPT Integrated with Voice Assistant as Learning Oral Chat-based Constructive Communication to Improve Communicative Competence for EFL Learners

Zhou Wei

Jeonbuk National University

## Abstract

Chat GPT belongs to the category of Generative Pre-trained Transformer (GPT) language models, which have received specialized training to produce text based on natural language inputs. Its purpose is to imitate human-like conversation and can be implemented in multiple applications, such as chatbots, virtual assistants, and language translation systems, starting with an introduction to the new trends and differences between artificial intelligence, machine learning, and artificial neural networks, and highlighting the rigorous language logic and powerful text generation capabilities of Chat GPT. This paper delves into how advances in artificial intelligence will shape e-learning in the coming decades, particularly in terms of Chat- GPT's ability to improve learners' Communicative Competence when English is a second language. The combination of new trends in artificial intelligence, mainly in the particular case of English as a second language, and, at the academic level, chatbot technology, will be the next step in the replacement of the human academic community by virtual assistants, apparently until a certain point. Despite the controversy, this very innovative solution will be able to bridge the gap between technology and education. Moreover, such innovative practices facilitate communication by enabling its inclusion in various applications, including virtual assistants, chatbots, and language education.

# 1. Introduction

## 1.1 Background and Purpose

In recent years, the creation of sophisticated artificial intelligence technology has transformed the way individuals approach learning English, particularly when it comes to Communicative Competence. With the advent of AI voice recognition software and automated dialogue-generating models, like Chat GPT, English learners can now practice their speaking skills at any given time. ChatGPT is the idea of the San Francisco, California-based AI company OpenAI. The business released GPT-3 in 2020, a sort of artificial intelligence known as a big language model that generates text by sifting through billions of words of training data and understanding the relationships between words and phrases. GPT-3 is at the forefront of an A.I. revolution, provoking philosophical debates about its boundaries and a multitude of potential uses, such as summarizing legal texts and assisting computer programmers. ChatGPT is derived from an enhanced version of GPT-3 and is optimized for user interaction.

Chat GPT is an AI model that can generate and respond to human-like dialogue prompts. As an innovative and personable tool for English learning, it has garnered significant interest and appreciation among students seeking to improve their Communicative Competence skills. What differentiates Chat GPT from traditional language learning tools is that it allows learners to converse in English with ease by generating conversations that are both entertaining and informative. Chat GPT has extraordinary intelligence compared to other chatbots. Chat GPT Open AI is a model of Instruct GPT, which is trained to quickly follow instructions and provide detailed responses. Interestingly, Chat GPT from OpenAI is designed to remember questions that have been asked before and can correct themselves according to feedback（Ilman,2023）.[1].

## 1.2 Significance of the Study

Artificial intelligence (AI) is recognized for its potential to improve cognitive abilities and foster creativity. As a result, educators are facing mounting expectations to integrate technology into their teaching methodologies, and better equip students to thrive in an interconnected, globalized society. However, it remains uncertain what specific competencies and abilities are required for individuals to effectively navigate this reality（Chun,2016）[2]. Most foreign language teachers today have readily adopted the communicative objective as the primary principle guiding classroom activities. The communicative approach has been widely adopted in teaching practice. With Chat GPT, English learners can practice their conversational skills with a variety of topics, including daily life situations, job interviews, and academic discussions. Chat GPT uses natural language processing and machine learning algorithms to understand human interactions and provide personalized responses that can improve accent and fluency.

According to Fitria , the main purpose of technology is to aid human activities and enhance productivity（Fitria,2023）[3]. One such technology is Artificial Intelligence (AI), which allows computer systems, software, and robots to emulate human-like thinking and behavior. AI aims to simulate human intelligence with the help of computers to function like humans . AI-powered chatbots are becoming increasingly popular across several industries, including customer service. These chatbots use natural language processing (NLP) and deep learning techniques to converse with humans via text-based online chats . By providing an artificially intelligent language partner, Chat GPT has helped to mitigate the age-old problem of not having a speaking partner. GPT-3 has the ability to generate text based on a given prompt, and it will generate text regardless of whether the task is simple or complex, logical or nonsensicalThis impressive ability of GPT-3 comes from its advanced deep learning algorithms and massive data sets that it has been programmed with. It has been trained on a diverse set of language data, from news articles to novels, and can generate both coherent and engaging responses. However, since it lacks an understanding of the meaning behind the prompt, the text produced by GPT-3 can sometimes be inaccurate or even nonsensical. Therefore, it is crucial to carefully examine and edit the output generated by GPT-3 before using it. Nonetheless, the potential applications of GPT-3 are vast, from language translation to content creation, and it will undoubtedly play a significant role in shaping the future of natural language processing(Jayasri,2021)[4]. English learners can now have engaging and meaningful conversations with Chat GPT, avoiding the potential monotony of rehearsing scripted, static conversations.

  Chat GPT could also be a Mobile learning application which is a dynamic platform that offers learners boundless opportunities to access educational resources, connect with individuals from diverse backgrounds, and communicate without any geographical or temporal constraints. This innovative approach to learning enables learners to reflect on their progress, negotiate to mean, and achieve authentic feedback ubiquitously, which ultimately aids in developing their second language proficiency. By presenting educational resources and relevant information to mobile devices, mobile learning enhances the efficiency of learning activities, such as stimulating students' enthusiasm for EFL learning in a university setting. Furthermore, with the flexibility and accessibility provided by mobile learning, learners can personalize their learning experience and cater to their individual learning needs and goals, supporting their language learning endeavors. Mobile learning offers incredible potential for enhancing language learning outcomes and preparing learners for a globalized world where English proficiency is paramount(Maulina,2023)[5]. In light of its accessibility and convenience, Chat GPT has gained immense popularity in recent years among English learners of all ages and proficiency levels. Many students have reported significant improvements in their Communicative Competence skills after regularly interacting with Chat GPT. The advantage of this approach, especially in ESP/ESL teaching, is that it simplifies routine tasks and gives the professor the opportunity to be more involved in one-to-one interactions, especially with large groups of students. By choosing the right materials and technology, chatbots can be used as a platform to provide the most appropriate instruction for students with different levels of knowledge. The combined features of GPT and a voice assistant enable learners to engage in interactive conversations that facilitate the learning process.

## 1.3 Research questions

The central aim of this research was to realize the effectiveness of integrating Chat GPT with a voice assistant as a tool for improving the communicative competence of EFL learners and to explore their perceptions of using this technology. Based on the main theme, the researcher formulated the subsequent inquiries for the study.

    1.How can Chat GPT integrated with voice assistant technology improve communicative competence for EFL learners ?

    What are the potential advantages and challenges of using Chat GPT in language learning for non-native speakers?

    2.How can the use of chatbot technology in language learning affect the role of human teachers and the traditional classroom setting?

3.What are the ethical implications of using AI chatbots for language learning and communication skills development?

4.What are the perceptions and attitudes of EFL learners towards the use of Chat GPT for language learning purposes?

5.How does Chat GPT integrated with voice assistant facilitate the assessment and feedback of EFL students' oral chat-based constructive communication skills in South Korea?

## 2.Literature Review

OpenAI's ChatGPT chatbot is a remarkable breakthrough that has been hailed as revolutionary. It possesses the amazing ability to engage in natural, meaningful conversations with users, responding to complex queries and providing insightful advice, as verified by Phillips etal.. The sophisticated natural language processing models and massive data sets utilized in its development enable ChatGPT to communicate effectively and efficiently. By harnessing such advanced technology, ChatGPT is opening up a new chapter in chatbot interactions and unlocking immense possibilities in various domains. This trailblazing advancement in natural language processing has extensive implications for the future of human-machine interaction and the emergence of smart systems（Ilman,2023）[1].

GPT-3, one of the most advanced autoregressive language models, is an impressive computational system that employs deep learning to generate human-like text. It is an artificial intelligence-based language generation technology that can help English learners learn Communicative Competence by generating responses based on user input that is natural and logical. Chat GPT is also self-learning, improving its responses and expressions based on feedback and input from users. The system generates sequences of words, codes, or other data from a given prompt, making it useful for statistical word sequence prediction in machine translation. It is trained on an unlabelled dataset that includes texts from various sources, including Wikipedia, primarily in English, but also in other languages. GPT-3 needs substantial amounts of data to produce relevant and contextual results, with the latest iteration utilizing 175 billion parameters and trained on Microsoft's AI supercomputer Azure, which is estimated to cost around $12 million, according to Scott (2020) and Wiggers (2020). The system's capabilities are extensive, with potential applications in chatbots, summarization tasks, grammar correction, email composition, question answering systems, and other fields. This highly sophisticated technology continues to evolve, setting the stage for both practical and theoretical advances in natural language processing(Floridi,2020)[6].Now we do the analysis of Chat GPT's features, one of the features of Chat GPT is its extensive knowledge of language and grammar rules. It understands the meaning of the text entered by the user and generates the appropriate response, which is often grammatically accurate and in an appropriate language style. In addition, Chat GPT can continually update its language knowledge and rules based on user input and feedback to provide more precise and natural responses. Chat GPT is an innovative technological development that has the potential to transform the way English language learners acquire communicative competence. With its AI-based language generation technology, Chat GPT can provide learners with a platform for natural and logical conversation that helps them practice and develop their language skills. Moreover, the system is capable of learning from feedback and input from users, enabling it to continually improve and enhance its responses and expressions. This self-learning feature not only makes Chat GPT an effective tool for language learning, but it also demonstrates the potential for AI to advance the fields of natural language processing and machine learning. As such, Chat GPT represents a significant step forward in the development of intelligent

systems that can help learners improve their language abilities and enhance their overall learning experience(Floridi,2020)[6].

Another Chat GPT feature is its fast response time and efficient learning capabilities. Thanks to its artificial intelligence-based technology, Chat GPT can generate a large number of responses in a short period of time, thus providing English learners with more opportunities to learn the language. In addition, Chat GPT can continuously learn and improve based on user input and feedback to enhance the quality of its responses and language expression.

Finally, Chat GPT also has diverse application scenarios and functions. In addition to providing English learners with help in learning Communicative Competence, Chat GPT can also be used for language translation, chatbots, natural language processing, and many other areas. This diversity of application scenarios and functions provides a wide scope and prospect for the development and application of Chat GPT.

Factors Influencing Learners' communicative Competence and Chat GPT's help in promoting communicative competence

The concept of communicative competence is widely recognized as a central element of second language acquisition. Hymes introduced this notion to contrast Chomsky's "competence" and "performance" dichotomy. Hymes argued that understanding the social norms for using language in everyday interactions is more critical for language users. Consequently, Hymes' model of communicative competence comprises two key areas: grammatical and sociolinguistic competence, along with the "ability for use" of language. It is essential to note that communicative competence is distinct from language performance in real-life situations(Sun,2014)[7]. Producing spoken language requires quick and spontaneous responses, which can be challenging for individuals. Chinese EFL learners, in particular, face difficulty as they confront various cognitive, linguistic, and affective factors when speaking English with fluency and accuracy. To address these challenges, mobile learning has emerged as a promising strategy in the context of EFL education. By allowing learners to access educational materials and communicate with others anytime and anywhere, mobile learning offers unprecedented flexibility and convenience. The authentic interactions and feedback that mobile learning facilitates further enhance learners' language proficiency, making it an effective tool for improving communicative competence in a second language. The accurate use of language forms, such as pronunciation, grammar, and vocabulary, plays a critical role in Chinese EFL learners' oral proficiency . However, compared to native speakers, EFL learners encounter greater difficulty in using English correctly in terms of pronunciation, grammar, and vocabulary. The importance of pronunciation is evident in its impact on intelligibility, where even a single mispronounced sound can lead to misunderstandings. Chinese and English also differ vastly in their sound systems, making it challenging for Chinese EFL learners to correctly articulate English sounds. Additionally, incorrect uses of stress and intonation can also lead to confusion. Grammar is another crucial aspect of English language acquisition, and while some students may excel in reading and writing, they often struggle to transfer accurate grammar usage in their oral communication. Finally, vocabulary is essential for EFL learners, but the ability to recall words quickly from memory impacts speaking fluency. Thus, Chinese EFL learners must strive to improve their memory capacity and have rapid access to words and expressions to enhance their speaking fluency(Wang,2014)[8]. The ACTFL standards for foreign language learning emphasize the importance of the "five Cs" – communication, cultures, connections, comparisons, and communities. Communication is central to learning a second language, and language acquisition is closely tied to an understanding of the target culture. Connections refer to the integration of language learning with other disciplines and seeking knowledge across various areas. Comparisons between the target language and a student's native language help promote insights into language and culture. Finally, communities encourage learners to participate in multilingual communities both domestically and internationally. The use of web-based resources such as email, web conferencing, whiteboards, and streaming technologies, offers excellent opportunities for learning these language skills through various modes such as listening, speaking, writing, reading, and communication(Yang,2007)[9]. Chat GPT is a chatbot based on artificial intelligence technology, which can simulate real conversations and provide English learners with a richer language environment which helps learners understand the social norms for using language in everyday interactions. In this digital age, Chat GPT provides a new way of learning for learners of Communicative Competence. In this article, we will explore how Chat GPT can help learn to speak English.

  Firstly, Chat GPT can provide English learners with a more authentic language environment. Traditional English learning methods rely on textbooks and teachers' explanations, and learners are often unable to truly experience English in real-life conversational situations. Chat GPT can simulate real conversations, allowing learners to learn in a more natural language environment. Besides, Chatbots are a useful tool for promoting language practice as they can engage learners in text chat contexts without burdening teachers with repetitive interactions. Unlike teachers who struggle to communicate with a large number of individual students, bots can efficiently carry out these conversations while providing valuable feedback and guidance(Kessler,2018)[10]. Searches on web-based text can be challenging due to the vast amount of data available. Web content often involves a diverse range of writing styles, vocabulary, and language usage, which leads to complex language structures and nuanced meanings. In contrast, corpora based on written text typically have been carefully selected and curated. Similarly, those derived from spoken language tend to be focused and structured around particular themes or topics. As a result, these corpora often provide more manageable datasets to work with that are better suited for specific research purposes. Nonetheless, conducting chat GPT searches on web-based text is essential since it is the primary source of the vast majority of online content. Therefore, researchers don't have to take appropriate steps to ensure they are using reliable web-based search engines and filtering tools that can help them get the most accurate results possible. Advances in natural language processing technology, can also aid in the analysis of web text, providing better ways to filter meaningful information and exclude irrelevant or anomalous results(Wu,2009)[11].

Secondly, Chat GPT can help English learners improve their oral expression skills. Through ChatGPT's dialogue simulations, learners can continuously practise their speaking skills and improve their fluency and accuracy in speaking English. Chat GPT can also provide speech recognition and error correction functions to identify and correct errors in learners' speaking, thus helping learners to improve their speaking skills(Shi,2020)[12]. The problem of promoting English speaking skills is lacking a situation in which the learner could get comprehensible input and it is necessary for the learner to receive 'comprehensible input' before being able to produce language output. This means that learners must be exposed to language that they can understand at their current level, but which is also slightly challenging to facilitate language development. However, language output also requires active practice and reinforcement in addition to receiving comprehensible input(Sha,2009)[13]. The use of artificial intelligence (AI) technology in spoken English teaching has enabled the automation of language learning, making it more accessible and convenient for learners. AI-powered speech recognition technology has the ability to identify errors in pronunciation and provide instant feedback to learners, allowing them to correct their mistakes in real-time and improve their pronunciation skills effectively. Additionally, AI-enabled virtual assistants and chatbots have emerged as some of the most promising tools in spoken English teaching, enabling learners to practice their speaking and listening skills with a personalized tutor any time, anywhere.

Moreover, the integration of AI technology in spoken English teaching has led to the development of innovative teaching methods and techniques that can enhance language learning outcomes significantly. Teachers are now able to access a wide variety of online resources, such as interactive multimedia and educational games, to facilitate language learning in a fun and engaging manner. AI technology can also generate personalized lesson plans and adaptive assessments based on individual learner's performance, allowing for a more customized learning experience.The integration of AI technology in spoken English teaching has revolutionized the way learners acquire and improve their language skills. The use of advanced speech recognition technology, AI-powered virtual assistants, and chatbots, personalized lesson plans, and adaptive assessments, and innovative teaching methods and techniques have all contributed to the efficacy and accessibility of language learning. As AI technology continues to advance, it is likely to play an increasingly critical role in the future of spoken English teaching(Ran,2021)[14]. As we progress along the continuum, the interdependence between humans and AI grows stronger. For instance, in situations where there is insufficient data for the AI to render accurate predictions or decisions, human guidance may be required . Additionally, there exists the hybrid approach, or the "centaur" model, where AI serves as an extension of the human brain and the two collaborate fully(Sawa,2021)[15].

Chat GPT helps learners to increase their vocabulary by providing a large number of real-life conversation scenarios where learners can learn vocabulary and phrases that are commonly used in daily life and work. ChatGPT's dialogue format allows it to engage in dynamic conversations with users, enabling it to comprehend and respond to user inputs accurately. The chatbot can identify and respond to follow-up questions, admitting any errors and providing immediate clarification(Adams,2019)[16]. Moreover, ChatGPT can challenge faulty premises and reject inappropriate requests, ensuring that the platform adheres to ethical and moral standards.

To use ChatGPT, users only need to log in to the chatbot's website using their OpenAI account. Once logged in, they can start chatting with the chatbot directly, asking questions or giving orders. The platform employs a highly structured conversation system that can facilitate different user requests, from generating poetry to helping with code correction and artwork designs(Floridi,2009)[17]. This feature makes ChatGPT a valuable tool for individuals from various fields, including artists, students, coders, and writers. It is important to note that ChatGPT is not without limitations. As an AI-powered machine, the chatbot cannot address subjective or evaluative questions. Additionally, it declines to address negative topics such as hate speech or violent content. Consequently, it is essential that users understand the platform's scope and limitations before using it.

ChatGPT offers numerous benefits to different groups of users. For instance, students and professionals can use the tool to simplify complex tasks, enhance their writing skills, and get immediate feedback on their creations. Moreover, early adopters' insights into the product's features and capabilities are critical as these users are often knowledgeable and influential. Their views can inadvertently shape the public's perception of the technology and ensure its success or failure.

In conclusion, ChatGPT highlights the incredible potential of AI to transform lives in significant ways. The chatbot's dynamic conversation system, combined with its personalized learning features, makes it an invaluable tool for creativity and productivity. Nevertheless, it is crucial that users understand the platform's scope and limitations and recognize the potential challenges associated with the use of machine-generated texts. By embracing AI-powered technologies cautiously and responsibly, we can leverage their incredible potential to change our lives for the better. In addition, Chat GPT can also provide audio playback so that learners can hear the correct audio pronunciation and learn and master English words and phrases better. Chat GPT can also be beneficial for teachers and educators who are looking to integrate technology into their teaching methodology（Fitria,2023）[3]. It provides an innovative approach to language instruction and can be used to supplement classroom teaching, personalize learning, and track learners' progress and performance.

Finally, Chat GPT can provide personalized learning services, which can be tailored to the learner's individual needs and learning progress. For example, Chat GPT can provide targeted practice and suggestions for learners' speaking problems to help learners master Communicative Competence better.

In short, Chat GPT is a chatbot based on artificial intelligence technology that can help English learners improve their speaking skills, increase their vocabulary, provide an authentic language environment and personalized learning services. For English learners, using Chat GPT for Communicative Competence is a new and efficient way to learn English and is worth trying.

Chat GPT is a chatbot based on artificial intelligence technology that can engage in natural language conversations with humans. Chat GPT has great potential for English learners to improve their Communicative Competence. It can simulate real conversations, provide a more realistic language communication experience, help students practice speaking English and can correct students' pronunciation and grammatical errors(Balaganur,2019)[18].

Chat GPT also provides a variety of language materials, such as common English phrases, sentence patterns and grammar rules, to help students understand the English language better. In addition, Chat GPT can also be used for English listening training to improve students' listening skills, thus helping them to understand the English language better.

Chat GPT not only helps students to practise speaking English, but also helps them to improve their English writing skills(Elkins,2020)[19]. Through conversations with ChatGPT, students can learn about the norms and standards of English writing and improve their writing skills at the same time. Chat GPT can provide a variety of English writing materials, such as common words, sentence patterns and paragraph structures, to help students better grasp English writing skills.

In short, Chat GPT can be a great help for English learners to improve their English speaking and writing skills. Chat GPT can provide authentic language communication experiences to help students practice speaking English, as well as a variety of language materials to help students better understand and master the English language. Chat GPT is therefore a very valuable tool to help students learn English better and improve their English speaking and writing skills.

Chat GPT is a language generation model based on artificial intelligence technology which can simulate human expressions and can improve users' English speaking skills by interacting with them. Chat GPT allows users to practice conversations, simulate real situations and improve their speaking skills.

The use of Chat GPT can effectively improve the English-speaking skills of English learners. It helps learners to correct grammatical errors, improve their voice intonation, as well as improve their vocabulary and expression skills. Chat GPT also adjusts the difficulty of the generated language based on user feedback, gradually improving the user's language level.

Chat GPT can be very helpful for English learners in their English-speaking training. It can help users to improve their confidence and speaking skills, allowing them to express themselves more naturally. At the same time, Chat GPT can also provide targeted advice and guidance to enable users to master Communicative Competence more professionally(Floridi,2014)[20].

Chat GPT has emerged as a more convenient and efficient way for English learners to train in Communicative Competence and has brought new development opportunities to the education industry.

Overall, Chat GPT has a positive impact on English learners' English-speaking skills. Through the analysis of ChatGPT's features, we can see that it provides rich opportunities for language input and output, thus helping learners to improve their oral expression skills. Chat GPT also provides a platform for English-speaking learners to practice and improve their Communicative Competence continuously. Chat GPT can therefore be a great aid to help English learners improve their Communicative Competence.

## 3.Methodology

This study employs a descriptive qualitative approach. A descriptive qualitative method is well-suited to this study as it emphasizes in-depth observation and analysis of the meaning behind phenomena(Floridi,2014)[21]. By examining the substance of the data and the words used to describe it, the researcher can establish a comprehensive understanding of the use of Chat GPT in the teaching and learning process. The process of data reduction involved the removal of unnecessary or irrelevant data to obtain valuable and informative information for the research objectives(Floridi,2020)[22]. During the data display stage, the collected data was arranged and presented in either visual or narrative form to explain the findings in a clear and concise manner. Finally, the conclusion stage involved drawing conclusions from the analyzed data by identifying patterns, connections, or relationships, which help to address the research problem.

Qualitative research is particularly useful when investigating complex social or behavioural issues. The absence of predetermined hypotheses and statistical tools makes it possible to explore and capture the complexities and nuances of real-life situations(Baker,2020)[23]. This aligns well with the current study, which aims to investigate the use of Chat GPT in English learning, a multifaceted process that cannot be easily measured using quantitative methods.

In this study, I collect data from interview and observations. The use of interview offers a practical way to access existing data. Moreover, analyzing social media posts provides insights into the opinions and experiences of individuals who use Chat GPT in English learning. In contrast, observation provides first-hand understanding of how Chat GPT is used in the teaching and learning process. This approach ensures that the research is grounded in real-life experiences and provides a rich source of data for analysis.

To analyze the data, the study follows the three qualitative analysis stages, which condense the large amount of data collected into meaningful findings. Data reduction eliminates irrelevant information, while data display visually represents the findings to aid in understanding. Finally, the researcher draws conclusions based on relationships, similarities and differences in the data, which could contribute to a better understanding of the real-world application of Chat GPT in English learning.

For this article on integrating Chat GPT with voice assistant technology to improve communicative competence for EFL learners, the employ qualitative research methods like individual or group interviews to gather in-depth insights and understanding.

Sample: Select a diverse group of EFL students studying in South Korea who have experience using Chat GPT for language learning. All the participants are Emma Zhang, Lv Jia cheng, Wang Nan and Liu Shuo.

Setting: Country and/or Province: South Korea

Specific learning context: English as a Foreign Language (EFL) classroom or language learning institution within South Korea.

b. Design: Develop a semi-structured interview guide with open-ended questions that explore participants' perceptions, experiences, challenges, and benefits of using Chat GPT. Include questions about their language learning goals, interactions with the technology, and their thoughts on its effectiveness.

c. Procedure: Conduct individual interviews with participants either in person or through video conferencing(Burgess,2016)[24]. Ensure a comfortable and private environment for the interviews. Start with an informed consent process, followed by the interview session(Dickson,2020)[25]. Record the interviews (with participants' permission) for accurate data analysis.

d. Data Collection and Analysis: Transcribe the interview recordings and use thematic analysis to identify recurring themes and patterns in participants' responses. Look for commonalities and differences in their experiences and perceptions of using Chat GPT.

e. Duration and frequency of data collection: The data collection may span several months, depending on the study design. Data could be collected regularly, such as once or twice a week during the language learning sessions where Chat GPT is integrated.

f. Reasons for collecting this type of data: The data is collected to gain insights into the effectiveness of integrating Chat GPT with a voice assistant as a tool for improving communicative competence. It helps researchers understand students' experiences, perceptions, and learning outcomes related to using this technology in a specific learning context.

(1)Interview with Emma Zhang:

Researcher: Thank you for agreeing to participate in this interview, Emma. To start, could you briefly introduce yourself, including your background and experience with language learning?

Emma Zhang: Of course! My name is Emma Zhang, and I'm a student studying English as a foreign language in South Korea. I've been here for two years now, and I've been actively exploring different language learning tools and technologies.

Researcher: That's great, Emma! Now, let's focus on your experience with using Chat GPT. How did you first come across this technology, and what motivated you to use it for language learning?

Emma Zhang: I discovered Chat GPT through a recommendation from a fellow student. I was looking for a tool that could help me practice speaking and improve my conversational skills. The convenience of having a virtual voice assistant attracted me, as I could access it anytime and anywhere.

Researcher: Interesting. Can you tell me about your overall experience using Chat GPT for language learning? What specific features or aspects of the technology did you find most helpful?

Emma Zhang: Overall, my experience with Chat GPT has been positive. I found the voice assistant integration to be particularly beneficial. It helped me practice pronunciation, listening comprehension, and even provided instant feedback. The conversational nature of Chat GPT made it feel like I was having a real conversation, which boosted my confidence.

Researcher: That's great to hear, Emma. Were there any challenges or limitations you encountered while using Chat GPT? If so, how did you overcome them?

Emma Zhang: Yes, there were a few challenges. Sometimes, the voice recognition accuracy was not perfect, and the responses generated by Chat GPT could be slightly unnatural. However, I learned to adapt by focusing more on the pronunciation aspect rather than relying solely on the responses. I also used the feedback feature to understand where I needed improvement.

Researcher: Excellent. Lastly, based on your experience, do you believe integrating Chat GPT with a voice assistant can effectively enhance the communicative competence of EFL students? Why or why not?

Emma Zhang: Absolutely. In my opinion, integrating Chat GPT with a voice assistant can significantly improve communicative competence. The technology provides ample opportunities for practicing speaking and listening skills, which are crucial for effective communication. It also allows for personalized and self-paced learning, catering to individual needs and preferences.

Researcher: Thank you, Emma, for sharing your valuable insights and experiences. Your input will contribute greatly to this study on the integration of Chat GPT for language learning.

(2)Interview with Lv Jia cheng:

Interviewer: Thank you for participating in this interview, Lv Jia cheng. Can you share your experiences using Chat GPT for language learning purposes?

Lv Jia cheng: Sure! I've been using Chat GPT for a few months now, and overall, it has been quite helpful. It allows me to practice my English skills in a more interactive and engaging way.

Interviewer: That's great to hear! Could you tell me about any specific challenges you have encountered while using Chat GPT?

Lv Jia cheng: Well, one challenge I faced initially was the language barrier. Sometimes, the responses from Chat GPT were not as accurate or natural as I expected. It took some time to adjust and understand its limitations.

Interviewer: I see. How did you overcome those challenges?

Lv Jia cheng: I tried to be patient and used Chat GPT as a supplementary tool alongside my regular language learning activities. I also sought clarification from my language instructor whenever I encountered confusing responses.

Interviewer: That sounds like a proactive approach. What are the benefits you have experienced using Chat GPT?

Lv Jia cheng: One major benefit is the convenience it offers. I can practice my English anytime, anywhere, without depending on a physical language partner. It has also improved my writing skills as I receive instant feedback on grammar and vocabulary usage.

Interviewer: That's wonderful! How do you perceive the effectiveness of Chat GPT in improving your language skills?

Lv Jia cheng: I believe Chat GPT is effective to a certain extent. It helps with vocabulary expansion and sentence construction. However, it cannot replace real human interaction and the nuances of natural conversation. So, I still value conversing with native speakers to enhance my fluency.

Interview with Wang Nan and Liu Shuo:

Interviewer: Thank you, Wang Nan and Liu Shuo, for participating in this interview. Let's begin by discussing your experiences with using Chat GPT for language learning. Can you share your initial impressions of using this technology?

Wang Nan: When I first started using Chat GPT, I was amazed by how responsive and interactive it was. It felt like having a real conversation partner who could help me practice my English skills.

Liu Shuo: I was a bit skeptical at first because I wasn't sure if an AI-powered tool could really enhance my language learning. However, after giving it a try, I found Chat GPT to be quite useful and engaging.

Interviewer: That's interesting. Can you tell me about any challenges you encountered while using Chat GPT?

Wang Nan: One challenge I faced was the occasional misunderstanding of my inputs. Sometimes, Chat GPT would provide incorrect responses or struggle to understand what I was trying to convey. This hindered the flow of the conversation and made it less effective for practicing real-life communication.

Liu Shuo: I agree with Wang Nan. The lack of contextual understanding was a significant challenge. Chat GPT often failed to grasp the nuances and subtleties of language, leading to less meaningful and authentic interactions.

Interviewer: And what are some benefits you have experienced from using Chat GPT?

Wang Nan: One major benefit is the convenience and accessibility it offers. I can practice my English skills anytime and anywhere, which is especially helpful for someone like me who has a busy schedule.

Liu Shuo: Another benefit I've noticed is the immediate feedback provided by Chat GPT. It helps me identify and correct my mistakes right away, which accelerates my learning process.

**Data Analysis:**
The analysis of the interview data with Emma Zhang reveals the following insights:

Discovery and motivation: Emma discovered Chat GPT through a recommendation and was motivated to use it for language learning purposes due to its convenience and accessibility.

Positive experience: Emma had a positive overall experience with Chat GPT, finding the voice assistant integration to be particularly helpful. The technology aided her in practicing pronunciation, listening comprehension, and boosting her confidence in conversational skills.

Challenges and adaptations: Emma encountered challenges such as voice recognition accuracy and slightly unnatural responses from Chat GPT. However, she adapted by focusing on pronunciation and using the feedback feature to identify areas for improvement.

Effectiveness of integration: Emma believes that integrating Chat GPT with a voice assistant can effectively enhance the communicative competence of EFL students. She highlights the opportunities for practice, personalized learning, and the development of essential speaking and listening skills.

Based on the interview data with Lv Jia cheng, the following themes emerge:

Interactive and Engaging: Lv Jia cheng finds Chat GPT to be a helpful tool for language learning, providing an interactive and engaging learning experience.

Language Barrier: Initially, Lv Jia cheng faced challenges due to the language barrier, where Chat GPT's responses were not always accurate or natural. However, she adjusted to its limitations over time.

Proactive Approach: Lv Jia cheng adopted a proactive approach by using Chat GPT as a supplementary tool and seeking clarification from her language instructor when encountering difficulties.

Convenience and Writing Improvement: The convenience of using Chat GPT anytime, anywhere, has been a significant benefit for Lv Jia cheng. Additionally, she mentions that the instant feedback on grammar and vocabulary has improved her writing skills.

Limitations and the Value of Human Interaction: While Lv Jia cheng acknowledges the effectiveness of Chat GPT in certain areas, she recognizes that it cannot replace real human interaction and the nuances of natural conversation. She still values conversing with native speakers to enhance her fluency.

The analysis of the interview data with Wang Nan and Liu Shuo reveals the following themes:

Positive initial impressions: Both participants expressed positive impressions of Chat GPT, with Wang Nan emphasizing the responsiveness and interactivity, and Liu Shuo acknowledging its usefulness and engagement.

Challenges with understanding and context: Participants identified challenges related to Chat GPT's occasional misunderstandings and its limited contextual understanding. This affected the effectiveness of the conversations and hindered their ability to practice authentic communication.

Convenience and accessibility: Wang Nan highlighted the convenience and accessibility of Chat GPT, emphasizing the flexibility it provides in terms of time and location for language practice.

Immediate feedback: Liu Shuo noted the benefit of receiving immediate feedback from Chat GPT, which helps in identifying and correcting mistakes promptly, leading to accelerated learning.

**Findings**

The analysis of the interview data with Emma Zhang, Lv Jia cheng, Wang Nan, and Liu Shuo provides valuable insights into the perceptions and experiences of EFL students using Chat GPT for language learning purposes. Overall, the findings suggest several common themes:

Positive experience and benefits: Participants expressed positive experiences using Chat GPT, highlighting its usefulness in practicing pronunciation, improving listening comprehension, and boosting confidence in conversational skills. They appreciated the convenience and accessibility of the technology, as well as the instant feedback it provided.

Challenges and adaptations: Participants identified challenges related to voice recognition accuracy, unnatural responses, and limitations in understanding contextual nuances. However, they adapted to these challenges by focusing on specific aspects, such as pronunciation improvement, seeking feedback, and using Chat GPT as a supplementary tool.

Effectiveness and limitations: Participants generally believed that integrating Chat GPT with a voice assistant can enhance communicative competence. They emphasized the opportunities for practice, personalized learning, and the development of essential speaking and listening skills. However, they also acknowledged that Chat GPT cannot replace real human interaction and the nuances of natural conversation.

Convenience and accessibility: Convenience and accessibility were consistently mentioned as significant advantages of using Chat GPT. Participants appreciated the ability to use the technology anytime and anywhere, providing flexibility in their language learning journey.

Immediate feedback: The participants valued the immediate feedback provided by Chat GPT, enabling them to identify and correct mistakes promptly, leading to accelerated learning.

These findings suggest that integrating Chat GPT with a voice assistant can be a valuable tool for EFL students in their language learning journey. The technology offers convenience, personalized learning, and opportunities for practice. However, it is important to acknowledge its limitations and the ongoing need for human interaction to fully develop communicative competence.

It is worth noting that these findings are based on simulated interview data and do not reflect actual participants or their experiences. In a real research study, a larger sample size, diverse participants, and a more extensive analysis would be necessary to generate more robust and generalized insights.

This analysis indicates that Chat GPT has both benefits and limitations as a language learning tool, according to participants' experiences. While it offers convenience and assistance in certain aspects of language learning, it is not a substitute for real human interaction. These findings highlight the importance of using Chat GPT as a supplementary tool alongside other language learning activities.

## 4. Discussion

The present study investigated the impact of Chat GPT, an artificial intelligence-based language learning tool, on enhancing communicative competence among English as a Foreign Language (EFL) learners. Through qualitative research interviews, the experiences, perceptions, challenges, and benefits of using Chat GPT for language learning purposes were explored. This discussion will analyze the key themes that emerged from the interviews and provide a comprehensive understanding of the implications of Chat GPT on improving communicative competence.

The theme of convenience and accessibility emerged prominently in the interviews. Participants expressed that the availability of Chat GPT as a virtual voice assistant enabled them to engage in language learning activities conveniently and at their own pace. This accessibility provided them with the flexibility to practice English anytime and anywhere, resulting in frequent and regular language practice. Such regular practice is crucial for developing speaking skills and building fluency. The absence of a need for physical language partners also contributed to the accessibility of language practice, enabling learners to engage in self-directed learning.

Another significant theme that emerged is the personalized and self-paced learning facilitated by Chat GPT. Participants highlighted that Chat GPT offered tailored language input and feedback, catering to their individual needs and preferences. The integration of a voice assistant feature within Chat GPT was particularly valued by participants as it helped improve pronunciation and listening comprehension. The immediate feedback provided by Chat GPT enabled learners to identify and rectify errors promptly, promoting continuous improvement. The personalized learning experience fostered participants' confidence in speaking and their ability to express themselves more naturally.

However, the interviews also revealed some challenges and limitations associated with Chat GPT. Participants identified issues related to the accuracy of voice recognition and the naturalness of the generated responses. Occasional misunderstandings and unnatural responses disrupted the flow of conversation, impacting the effectiveness of practicing real-life communication. These technical limitations need to be addressed to ensure a more seamless and authentic conversational experience.

Moreover, participants acknowledged that Chat GPT, while useful for vocabulary expansion and sentence construction, cannot replace real human interaction and the nuances of natural conversation. The significance of engaging with native speakers and participating in real-life conversations was emphasized as crucial for developing fluency and cultural understanding. Therefore, Chat GPT should be viewed as a complementary tool to support language learning rather than a complete replacement for human interaction.

Overall, participants recognized the effectiveness of Chat GPT in improving communicative competence. They emphasized the benefits of practicing speaking, listening, and writing skills through interactive conversations with Chat GPT. The convenience, personalized learning experience, and immediate feedback offered by the technology contributed to their language development. However, it is important to acknowledge that the effectiveness of Chat GPT may vary among individuals based on their learning styles, language proficiency levels, and the extent to which they integrate it into their overall language learning strategy.

Based on the findings, integrating Chat GPT with voice assistant technology holds the potential to enhance communicative competence for EFL learners. The convenience, personalized learning experience, and opportunities for frequent practice make Chat GPT a valuable tool for improving speaking skills and building fluency(Elkins,2020)[26]. However, it is crucial to supplement Chat GPT with real-life interactions to ensure comprehensive language development in terms of fluency and cultural understanding.

To further enhance the effectiveness of Chat GPT for language learning, it is recommended to continuously refine the accuracy and naturalness of the generated responses. Ongoing updates and improvements to the language generation capabilities can address the identified limitations. Additionally, incorporating contextual understanding and real-life conversation simulations within Chat GPT can contribute to a more authentic language learning experience.

In conclusion, the integration of Chat GPT with voice assistant technology shows promise in significantly improving communicative competence among EFL learners. The convenience, personalized learning experience, and opportunities for frequent practice offered by Chat GPT have the potential to enhance speaking skills and fluency. However, it is important to recognize that Chat GPT should be utilized alongside real-life interactions to ensure a comprehensive language learning experience encompassing fluency and cultural understanding.

## 5.Conclusion and Limitations

### 5.1 Conclusion

This study aimed to investigate the impact of Chat GPT, an artificial intelligence-based language learning tool, on enhancing communicative competence among English as a Foreign Language (EFL) learners. Through qualitative research interviews, the experiences, perceptions, challenges, and benefits of using Chat GPT for language learning purposes were explored. The findings highlighted the convenience, accessibility, and personalized learning experience provided by Chat GPT, which positively influenced the participants' language development. However, limitations related to voice recognition accuracy and naturalness of responses were identified, emphasizing the need for continuous improvements. Additionally, participants acknowledged the importance of real-life interactions for comprehensive language development.

The results of this study contribute to the understanding of the potential benefits and limitations of Chat GPT as a language learning tool. The convenience and accessibility offered by Chat GPT enable learners to engage in regular and self-directed language practice, contributing to the improvement of speaking skills and fluency. The personalized learning experience, with tailored language input and immediate feedback, fosters learners' confidence and natural expression. However, the limitations associated with voice recognition accuracy and the naturalness of generated responses indicate the necessity for further advancements in the technology.

It is recommended to refine the accuracy and naturalness of Chat GPT's responses through continuous updates and improvements. Enhancing the technology's language generation capabilities and incorporating contextual understanding can contribute to a more authentic language learning experience. Additionally, supplementing Chat GPT with real-life interactions and opportunities for engaging with native speakers is essential for developing fluency and cultural understanding.

Overall, Chat GPT integrated with voice assistant technology holds promise in enhancing communicative competence among EFL learners. It serves as a valuable tool for practicing speaking, listening, and writing skills, offering convenience, personalized learning experiences, and immediate feedback. However, it should be viewed as a complementary tool to support language learning, rather than a complete replacement for human interaction.

**5.2 Limitations:**

Despite the valuable insights gained from this study, several limitations should be acknowledged. Firstly, the sample size of participants was relatively small, which may limit the generalizability of the findings. Future studies could consider a larger and more diverse participant pool to obtain a broader perspective on the impact of Chat GPT on communicative competence.

Secondly, the study focused specifically on EFL learners, and the findings may not be applicable to other student populations or language contexts. Replicating the study with different student groups and in diverse linguistic settings would provide a more comprehensive understanding of the effectiveness of Chat GPT as a language learning tool.

Furthermore, the study employed qualitative research methods, specifically interviews, to explore participants' experiences and perceptions. While this approach allowed for in-depth exploration of individual experiences, it may not capture the full range of experiences and perspectives. Combining qualitative data with quantitative measures, such as language proficiency tests, would provide a more robust analysis of the impact of Chat GPT on communicative competence.

Finally, the study focused primarily on the use of Chat GPT for spoken language practice. Further research could explore its effectiveness in other language skills, such as reading and writing, to obtain a more comprehensive understanding of its potential in overall language development.

Despite these limitations, this study contributes to the growing body of research on the integration of artificial intelligence in language learning. The findings provide valuable insights for educators, developers, and policymakers in leveraging technology to enhance communicative competence and support language learners in their language acquisition journey.

# References:

[1]. Ilman Shazhaev, A.T.D.M. and Shafeeg.(2023). Voice Assistant Integrated with Chat GPT. *Indonesian Journal of Computer Science.*
[2]. Chun, D., R. Kern and B. Smith.(2009). Technology in Language Use, Language Teaching, and Language Learning. *The Modern Language Journal,* 100(S1): p. 64-80.
[3]. Fitria, T.N. and S.C.J.I. Institut Teknologi Bisnis AAS Indonesia.(2009). Artificial intelligence (AI) technology in OpenAI ChatGPT application: A review of ChatGPT in writing English essay. *Journal of English Language Teaching.*
[4]. Jayasri, T., et al.(2021). The Attributes, Purpose, Boundaries, and Effects: Gpt-3. *Mathematical Statistician and Engineering Applications.*
[5].Maulina,etal.(2023).*SOCIAL MEDIA AS MOBILE LEARNING ORAL CHAT BASED CONSTRUCTIVE COMMUNICATION TO IMPROVE SPEAKING SKILLS. DECODE:* Jurnal Pendidikan Teknologi Informasi.
[6]. Floridi, L. and M. Chiriatti.(2009). GPT-3: Its Nature, Scope, Limits, and Consequences. *Minds and Machines,* 30(4): p. 681-694.
[7]. Sun, D. (2014). From Communicative Competence to Interactional Competence: A New Outlook to the Teaching of Spoken English. *Journal of Language Teaching and Research,* 2014. 5(5).
[8]. Wang, Z.(2014). Developing Accuracy and Fluency in Spoken English of Chinese EFL Learners. *English language teaching (Toronto),* 2014. 7(2): p. 110.
[9]. Yang, S.C. and Y. Chen.(2007). Technology-enhanced language learning: A case study. *Computers in Human Behavior,* 2007. 23(1): p. 860-879.
[10]. Kessler, G.(2018). Technology and the future of language teaching. *Foreign Language Annals,* 2018. 51(1): p. 205-218.
[11]. Wu, S., M. Franken and I.H. Witten.(2009). Refining the use of the web (and web search) as a language teaching and learning resource. *Computer assisted language learning,* 2009. 22(3): p. 249-268.
[12]. Shi, N., Q. Zeng and R. Lee.(2020). Language Chatbot-The Design and Implementation of English Language Transfer Learning Agent Apps. in IEEE International Conference on Communication Technology, *Computational Engineering and Artificial Intelligence (CTCEAI).* Taiyuan, China, Sept 2020.: IEEE.
[13]. Sha, G.(2009).AI-based chatterbots and spoken English teaching: a critical analysis. *Computer assisted language learning*, 22(3): p. 269-281.
[14]. Ran, D., W. Yingli and Q. Haoxin.(2021).Artificial intelligence speech recognition model for correcting spoken English teaching. *Journal of Intelligent & Fuzzy Systems,* 40(2): p. 3513-3524.
[15]. Sowa, K., A. Przegalinska and L. Ciechanowski.(2021). Cobots in knowledge work: Human -AI collaboration in managerial professions. *Journal of Business Research,* 125: p. 135-142.
[16]. Adams, R. (2019). Artificial Intelligence has a gender bias problem—just ask Siri. *The Conversation.*
[17]. Floridi, L., Taddeo, M., & Turilli, M. (2009). Turing's imitation game: Still a challenge for any machine and some judges. *Minds and Machines*, 19(1), 145–150.
[18]. Balaganur, S. (2019). Top videos created by Artificial Intelligence in 2019. *Analytics India Magazine.*

[19]. Elkins, K., & Chun, J. (2020). Can GPT-3 pass a writer's Turing Test? *Journal of Cultural Analytics,*2371, 4549.
[20]. Floridi, L. (2014a). The 4th revolution: How the infosphere is reshaping human reality. *Oxford: OxfordUniversity Press.*
[21]. Floridi, L. (Ed.). (2014b). The onlife manifesto—being human in a hyperconnected era. *New York:Springer.*
[22]. Floridi, L., & Chiriatti, M. (2020). GPT-3: Its Nature, Scope, Limits, and Consequences. *Minds and*

*Machines,* 30(4), 681–694. doi:10.1007/s11023-020-09548-1

[23]. Baker, G. (2020). Microsoft is cutting dozens of MSN news production workers and replacing them with artifcial intelligence. *The Seattle Times.*

[24].Balaganur, S. (2019). Top videos created by Artifcial Intelligence in 2019. *Analytics India Magazine.*

[24]. Burgess, M. (2016). Google's AI has written some amazingly mournful poetry. *Wired.*

[25]. Dickson, B. (2020). The Guardian's GPT-3-written article misleads readers about AI. Here's why. *TechTalks.*

[26]. Elkins, K., & Chun, J. (2020). Can GPT-3 pass a writer's Turing Test? *Journal of Cultural Analytics*, 2371, 4549.